\documentclass[twocolumn,showpacs,preprintnumbers,amsmath,amssymb]{revtex4}
\usepackage{graphicx}
\usepackage{dcolumn}
\usepackage{bm}

\def\ube13{UBe$\rm_{13}$}

\def\bi2212{Bi$\rm_2$Sr$\rm_2$CaCu$\rm_2$O$\rm_8$}
\def\ybi2212{Bi$\rm_2$Sr$\rm_2$YCu$\rm_2$O$\rm_8$}
\def\ycabi2212{Bi$\rm_2$Sr$\rm_2$Ca$\rm_{1-x}$Y$\rm_x$Cu$\rm_2$O$\rm_{8+\delta}$}

\def\y65cabi2212{Bi$\rm_2$Sr$\rm_2$Ca$\rm_{0.35}$Y$\rm_{0.65}$Cu$\rm_2$O$\rm_{8+\delta}$}
\def\Ir{CeIrIn$_5$}
\def\Co{CeCoIn$_5$}

\begin{document}

\preprint {\noindent submitted to Physical Review Letters  \hfill LA-UR-04-2278}

\title{Non-Fermi Liquid behavior in CeIrIn$_5$ near a metamagnetic transition.}

\author{C.Capan,$^1$ A. Bianchi,$^1$ F. Ronning,$^1$ A. Lacerda,$^2$ J.~D.~Thompson$^1$, P. G. Pagliuso,$^3$ J. L. Sarrao$^1$, and R. Movshovich,$^1$ }
\affiliation{$^1$Los Alamos National Laboratory, Los Alamos, New Mexico 87545\\
             $^2$National High Magnetic Field Laboratory, Los Alamos, New Mexico 87545\\
             $^3$Instituto de Fisica Gleb Wataghin, UNICAMP, 13083-970, Campinas, Brazil}

\date{\today}

\begin{abstract}

We present specific heat and resistivity study of \Ir\ in magnetic fields up to 17 T and
temperature down to 50 mK. Both quantities were measured with the magnetic field parallel to the
c-axis ($H \parallel$ [001]) and within the a-b plane ($H \perp$ [001]). Non-Fermi-liquid (NFL)
behavior develops above 12 T for $H \parallel$ [001]. The Fermi liquid state is much more robust
for $H \perp$ [001] and is suppressed only moderately at the highest applied field. Based on the
observed trends and the proximity to a metamagnetic phase transition, which exists at fields above
25 T for $H
\parallel$ [001], we suggest that the observed NFL behavior in \Ir\ is a consequence of a
metamagnetic quantum critical point.

\end{abstract}

\pacs{74.70.Tx, 71.27.+a, 74.25.Fy, 75.40.Cx}

\maketitle

Investigations of the material properties near a zero-temperature phase transition (Quantum
Critical Point or QCP) is at present a very active area of research, attracting both experimental
and theoretical attention. It is common for metallic compounds in the vicinity of a QCP to display
a variety of physical properties at odds with those expected for a Fermi Liquid (FL), a concept
that forms the basis for our understanding of the physics of a vast majority of metals.
Characteristic of a FL are such properties as a linear-in-temperature specific heat $C$ and
T-squared resistivity $\rho$. In contrast, materials near QCPs often display a diverging Sommerfeld
coefficient $\gamma = C/T$ and a power law temperature dependent resistivity $\rho = \rho_0 + A
T^\alpha$, with $\alpha$ significantly different from 2~\cite{stewart:rmp-2001}. The theoretical
picture of a system near a QCP is not complete at the moment, and the origins of the behavior
described above are the subject of intense theoretical investigations.

For a large number of the material studied, the two competing phases at the QCPs are
antiferromagnetically (AF) ordered and paramagnetic ones. It was demonstrated that for this class
of compounds pressure was an effective parameter for tuning a system through an AF QCP,
particularly for the Ce-based heavy-fermion compounds. This group of materials include, for
example, CeCu$_2$Ge$_2$~\cite{jaccard:pla-92}, CeRh$_2$Si$_2$~\cite{movshovich:prb-96},
CePd$_2$Si$_2$, and CeIn$_3$~\cite{mathur:nature-98}. Alternatively, in an increasing number of
materials, i.e. CeCu$_{5.2}$Ag$_{0.8}$\cite{heuser:physicaB-99b},
CeCoIn$_5$~\cite{bianchi:prl-03b,paglione:prl-03}, and
YbRh$_2$Si$_2$\cite{gegenwart:prl-02,gegenwart:jltp-03}, it was found to be possible to tune the
system through a QCP by varying the magnetic field.

A novel route to quantum criticality was proposed based on Sr$_3$Ru$_2$O$_7$, when its resistivity
in magnetic field close to the metamagnetic field $H_M = 8$ T displayed pronounced NFL
behavior~\cite{perry:prl-01,grigera:science-01}. On general grounds a first order $T = 0$ phase
transition does not lead to a QCP. Nevertheless, it was suggested that quantum critical behavior
can be associated with a first order phase transition when the transition's critical end point is
driven to zero temperature~\cite{grigera:science-01,millis:prl-02}. In Sr$_3$Ru$_2$O$_7$ the
critical end point of the metamagnetic phase transition temperature $T_M^c$ can be tuned by varying
the direction of the magnetic field with respect to the tetragonal crystal lattice. When magnetic
field is close to $H
\parallel$ [001], $T_M^c$ is suppressed close to zero, leading to the quantum critical behavior
observed in Sr$_3$Ru$_2$O$_7$~\cite{grigera:science-01}. Recent analysis indicate that
CeRu$_2$Si$_2$ may also be close to a metamagnetic QCP~\cite{flouquet:physicaB-02}.

Related phenomena may be in play in \Ir. For most of the compounds with AF QCPs magnetic field
suppresses the AF state, with the Fermi Liquid behavior recovered in the paramagnetic state above
the critical field of the QCP. In this Letter we present results of the specific heat and
resistivity measurements in \Ir\ which show the reverse trend, with magnetic field suppressing
rather than enhancing the Fermi Liquid state, pointing perhaps to a different route to Quantum
Criticality. Based on our results we suggest that the NFL behavior in \Ir\ is due to the proximity
to a metamagnetic phase transition and a metamagnetic QCP, perhaps similar to the recently
discussed case of Sr$_3$Ru$_2$O$_7$~\cite{grigera:science-01}.

The details of sample growth and characterization are described in
Ref.~\onlinecite{petrovic:epl-01}. Large  plate-like single crystals, up to 1 cm long, are grown
from an excess In flux. \Ir\ is a layered tetragonal heavy fermion compound from the 1-1-5 family,
with no long-range magnetic order but a superconducting ground state~\cite{petrovic:epl-01}. The
presence of the cylindrical Fermi surface sheet, inferred from de Haas-van Alphen
studies~\cite{haga:prb-01}, and a ratio of 4.8 of the effective masses between the c-axis and the
CeIn$_3$ planes~\cite{shishido:jpsj-02} makes \Ir\ a moderately anisotropic system. Moreover, the
crystal electric field effects result in an anisotropic susceptibility, with a step-like feature
around 50 K along the c-axis~\cite{pagliuso:physicaB-02}. The anisotropy of the spin fluctuations
is evidenced in the temperature dependence of $T_1$ derived from the NQR data~\cite{zheng:prl-01}.
This anisotropy is reflected in both the specific heat and resistivity data shown below.


\begin{figure}
\includegraphics[width=3.5in]{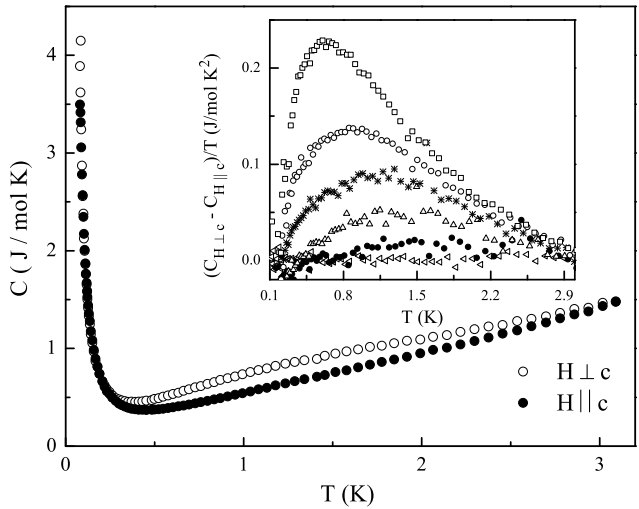}
\caption{\label{Anisotropy-hc-fig1} Specific heat as a function of temperature at 17 T for field
oriented ($\bullet$) parallel to c-axis  and ($\circ$) in the plane. Inset: Anisotropy of specific
heat (difference of in-plane and out-of-plane specific heat) as a function of temperature at
various magnetic fields up to 17 T. ($\triangleleft$) 1 T, ($\bullet$) 3 T, ($\triangle$) 6 T,
($\ast$) 9 T, ($\circ$) 12 T, ($\Box$) 17 T. }
\end{figure}

We measured the specific heat of a CeIrIn$_5$  single crystal with the quasi-adiabatic heat pulse
method in a dilution refrigerator between 100 mK and 3 K and in magnetic fields up to 17 T.
Fig.~\ref{Anisotropy-hc-fig1} shows the specific heat as a function of temperature in the normal
state, for the magnetic field of 17 T applied perpendicular and parallel to the c-axis. At low
temperature (below 200 mK) the specific heat is dominated by the nuclear Schottky anomaly, which is
mainly due to In nuclear levels split by the magnetic field. The Schottky anomaly can be well
approximated with an $\alpha/T^2$ dependence in the whole field range, with $\alpha \propto H^2$
for both field orientations, as expected. The 17 T field induces a significant shift between the
in-plane and out-of-plane specific heat, for temperatures ranging from 0.2 K to 3 K.


\begin{figure}
\includegraphics[width=3.5in]{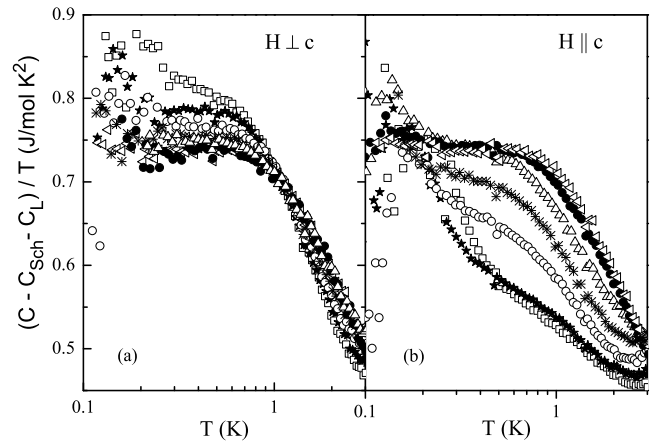}
\caption{\label{GammaElectronic}Electronic specific heat, obtained by subtracting the Schottky and
lattice contributions from the total specific heat, as a function of temperature at various
magnetic fields up to 17 T with the field applied (a) in plane  and (b) along the c-axis.
($\triangleleft$) 1 T, ($\bullet$) 3 T, ($\triangle$) 6 T, ($\ast$) 9 T, ($\circ$) 12 T, ($\star$)
15 T, ($\Box$) 17 T.}
\end{figure}


Figure~\ref{GammaElectronic} shows the electronic specific heat as a function of temperature on a
semi-log scale, for magnetic fields oriented (a) in-plane and (b) out-of-plane, with field values
ranging between 1 T and 17 T. The electronic specific heat is obtained after subtracting the
lattice and the nuclear Schottky contribution from the measured specific heat. The lattice
contribution is small in the temperature range of interest (only 2.8\% of the total specific heat
at 3K) and has been calculated from the LaIrIn$_5$ specific heat in the Debye approximation.

The Sommerfeld coefficient $\gamma = C/T$ rises as the temperature is reduced below 3 K, consistent
with earlier reports on non-Fermi-liquid behavior in this compound, both of specific
heat~\cite{kim:prb-02} and thermal expansion~\cite{oeschler:prl-03}, and reaches a plateau below
about 1 K in low magnetic fields. The saturation of $\gamma$ marks the onset of the Fermi liquid
regime below the temperature $T_{FL}$ for both field orientations. However, $\gamma$ has a
remarkably different evolution for the two field orientations studied when the magnetic field is
increased. Namely, the field in-plane does not have a strong effect on the overall shape of
$\gamma$, and just slightly suppresses $T_{FL}$ to lower temperatures and makes the slope above 1 K
steeper. In contrast, for $H
\parallel c$ the knee in $\gamma$ gradually disappears, and the overall slope becomes more flat
with increasing field, leading to the gap between the bare specific heat curves described earlier
(see Fig.~\ref{Anisotropy-hc-fig1}). Note that the difference in the slopes above 1 K cannot be due
to the error in subtraction of the Schottky contribution, which drops to $\approx 50$\% of the
total specific heat at 300 mK for 17 T.

The difference in the evolution of the specific heat with the field in different orientations is
emphasized in the inset of Fig.~\ref{Anisotropy-hc-fig1}, where we plot the difference between the
in-plane and out-of plane specific heat divided by temperature, $(C_{H\perp c}-C_{H
\parallel c})/T$. A broad maximum is resolved above the field of 3 T, reflecting the suppression of the specific heat in the $H
\parallel$ [001] orientation. This maximum increases in magnitude and shifts to lower temperatures
as the field is increased, reaching 0.23 $\rm J/mol K^2$ at 17 T around 0.6 K, or about 27\% of the
total specific heat.

There is a clear anisotropy in the evolution of $T_{FL}$ as well. At 1 T the Fermi liquid behavior
survives up to $\sim 0.9$ K independent of the field orientation. However, as the magnetic field is
increased, the Fermi temperature is depressed much faster when the field is along the c-axis, in
contrast to a very gradual decrease observed when the field is in-plane. The knee eventually
vanishes completely and $\gamma$ becomes divergent down to the lowest temperatures measured for
fields above 12 T with $H \parallel c$. This NFL behavior suggests that \Ir\ may be approaching a
QCP for fields $H \parallel c$ above 12 T.


\begin{figure}
\includegraphics[width=3.5in]{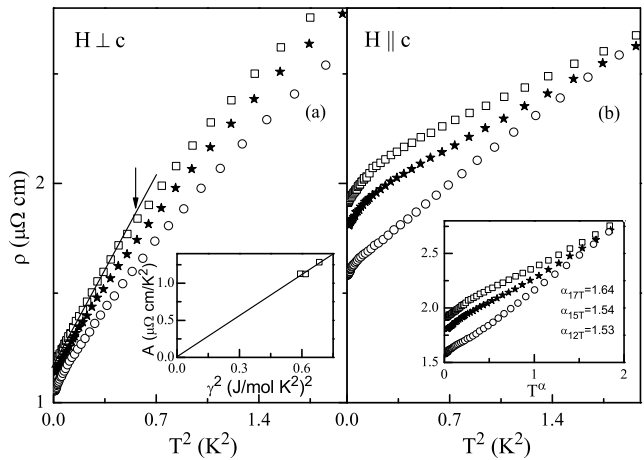}
\caption{Resistivity of \Ir\ at ($\circ$) 12 T, ($\star$) 15 T, and ($\Box$) 17 T. (a): $H
\parallel$ [001] orientation. Arrow indicate the onset the deviation of the 17 T data from the
Fermi Liquid $\rho \propto T^2$ behavior (straight line) at $T_{FL}$. Inset: coefficient $A$ of the
$T^2$ term in resistivity versus square of the Sommerfeld coefficient of $\gamma = C/T$ for $H
\perp$ [001], testing the Kadowaki-Woods relation (see the text). (b): $H \perp$ [001] orientation.
Inset: resistivity for $H \parallel$ [001] plotted vs. $T^\alpha$ for best fit of the lowest
temperature data. $\alpha \ne 2$ for these fields.} \label{resistivity}
\end{figure}

Figure~\ref{resistivity} shows resistivity data for magnetic fields of 12 T, 15 T, and 17 T applied
both within and out of the a-b plane of \Ir. Fermi liquid $\rho = \rho_0 + A T^2$ behavior is
clearly obeyed by the resistivity for all fields $H \perp$ [001] (Fig.~\ref{resistivity}(a)) below
a well defined temperature, marked by the arrow for 17 T data as an example. This behavior of the
resistivity is consistent with the FL behavior displayed at low temperature by the specific heat
for $H \perp$ [001] (see Fig.~\ref{GammaElectronic}(a)). To analyze the data further we plot
coefficient $A$ of the $T^2$ term in resistivity versus $\gamma^2$ in the inset of
Fig.~\ref{resistivity}(a). For many heavy fermion compounds $A$ vs $\gamma^2$ points fall close to
a single straight line~\cite{kadowaki:ssc-86}. Similar behavior is observed for \Ir, with $A
\propto \gamma^2$, as is seen in the inset. The coefficient of proportionality (Kadowaki-Woods
ratio) is about a factor of five smaller than the average value for other heavy-fermions. This,
however, is within the scatter displayed by this class of materials. Therefore, both resistivity
and specific heat of \Ir\ have a well developed heavy-fermion FL ground state for $H\perp c$ below
17 T.

Figure~\ref{resistivity}(b) shows resistivity data for $H \parallel$ [001]. In this case
resistivity above 12 T has a clear curvature at low temperature when plotted against $T^2$. In the
inset of Fig.~\ref{resistivity}(a) we plot resistivity versus $T^\alpha$, where $\alpha$ is
adjusted for the best fit to the data in the low temperature regime. The values of $\alpha$
obtained in such fashion are displayed in the inset, and are close to 1.5. For all three fields the
data plotted vs. $T^\alpha$ falls on a straight line below $\approx 0.5$ K. Such NFL behavior of
resistivity with the power law exponent different from 2 was observed in other compounds near a
QCP~\cite{stewart:rmp-2001}. Once again, the NFL resistivity behavior is consistent with the NFL
behavior displayed by the specific heat of \Ir\ for $H
\parallel$ [001]. Together, resistivity and specific heat data point to an approaching QCP with
magnetic field $H \parallel$ [001] increasing above 12 T.


The NFL behavior in \Ir\ is qualitatively different from that commonly observed in many heavy
fermion compounds in the vicinity of the QCP, where application of the magnetic field suppresses
the magnetically ordered state and stabilizes the FL behavior for $H > H_{QCP}$. In contrast, in
\Ir\ increasing magnetic field suppresses the FL state. Magnetic field in the range investigated
here, therefore, drives the system closer to a QCP. Such behavior is likely related to the
properties discovered in the very high field investigations of \Ir. Magnetization studies of \Ir\
revealed a metamagnetic anomaly at a field of 42 T~\cite{takeuchi:jpsj-01} at 1.3 K. Subsequent
specific heat measurements of \Ir\ uncovered  a phase transition into a magnetic state between 35 T
at 1.8 K and 45 T at 4.2 K, extrapolating to $H_M \approx 26$ T at zero
temperature~\cite{kim:prb-02}. More recent data from cantelever magnetometry investigation of \Ir\
show an anomaly in magnetic response at 30 T at 45 mK~\cite{palm:physicaB-03}. Despite the
quantitative differences, these measurement are most likely studying the same phenomenon of
metamagnetism in \Ir.

\begin{figure}
\includegraphics[width=3.5in]{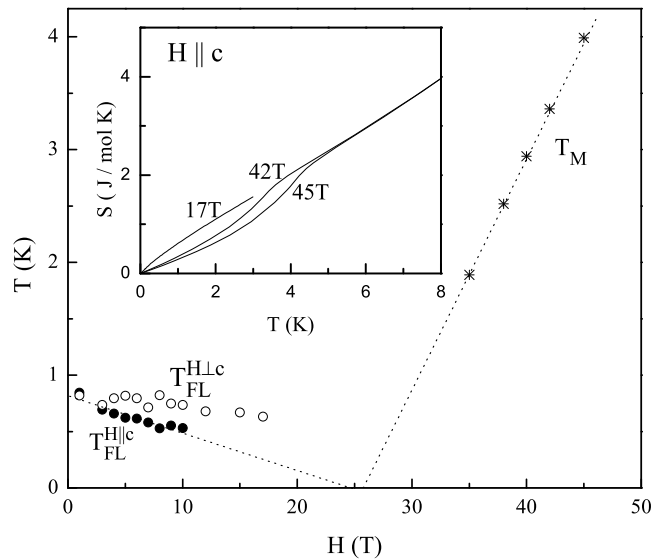}
\caption{H-T phase diagram of CeIrIn$_5$. Fermi Liquid temperatures $T_{FL}$ are derived from the
electronic specific heat are represented for the field in-plane ($\circ$) and parallel to c-axis
($\bullet$). The metamagnetic transition temperature $T_M$ is from Ref.~\cite{kim:prb-02}. Inset:
the entropy for our 17 T data and the 45 T data of Ref.~\cite{kim:prb-02}. See text for the details
on the extrapolations of both sets of data.} \label{HTphasediagram}
\end{figure}

It is instructive to compare behavior of \Ir\ with that of Sr$_3$Ru$_2$O$_7$, a material whose NFL
behavior close to a metamagnetic phase transition have recently received both experimental and
theoretical attention~\cite{grigera:science-01,millis:prl-02}. The Sommerfeld coefficient $\gamma$
of Sr$_3$Ru$_2$O$_7$ diverges in magnetic field close to the metamagnetic field $H_M = 8 T$ for $H
\parallel$ [001], close to the orientation when the first order critical end point is suppressed to
$T = 0$~\cite{zhou:prb-04}. It is interesting to note that even when the critical end point
temperature $T_M^c$ is finite (i.e. $H \perp$ [001], where $H_M \approx 6$ T) the phase diagram
showing NFL behavior, based on the resistivity data, is remarkably similar to the $H\parallel$
[001] case with $T_M^c = 0$~\cite{perry:prl-01}. In addition, $\gamma$ in Sr$_3$Ru$_2$O$_7$ is
close to being logarithmically divergent with temperature for $H \perp$ [001] at 6 T as
well~\cite{capan:unpublished}. Finally, we note that MnSi provides evidence that a first order
phase transition does not preclude NFL behavior~\cite{doiron-leyraud:nature-03}.

The above comparison leads us to suggest that the NFL behavior in \Ir\ is due to the proximity to a
metamagnetic phase transition. Figure~\ref{HTphasediagram} shows the magnetic field - temperature
phase diagram. The open and solid circles represent the Fermi Liquid temperature $T_{FL}$ obtained
from the analysis of the specific heat data for $H\perp c$ and $H\parallel c$, respectively. The
data for $H
\parallel c$ extrapolates to a value close to 25 T. Stars, representing the metamagnetic transition
temperature $T_M$ from Ref.~\cite{kim:prb-02}, extrapolate to about 26 T. The two values are very
close and hint at the possibility that the NFL behavior in \Ir\ for $H \parallel c$ between 12 and
17 T may be related to the metamagnetic phase transition observed at higher fields. To test this
hypothesis further we compared the entropy associated with the metamagnetic transition at 45 T with
the entropy of \Ir\ at 17 T.  We estimated the specific heat of \Co\ at 42 T and 45 T below 1.2 K
by linearly extrapolating $\gamma$ from Ref.~\cite{kim:prb-02} from 1.2 K to $T = 0$. The
extrapolated curves were integrated and the resulting entropy curves are displayed in the inset of
Fig.~\ref{HTphasediagram}. The entropy values are very close at high temperature above the
metamagnetic anomaly in specific heat. This indicates that the metamagnetic anomaly is built out of
the spin fluctuations that lead to the NFL behavior of specific heat of \Ir\ at 17 T, and presents
a strong argument in favor of the metamagnetic quantum critical point in \Ir\ being the origin of
the NFL properties of \Ir\ we observed.

For $H\perp c$, the metamagnetic transition does not occur below 52 T~\cite{takeuchi:jpsj-01}. This
is reflected in our data, with  FL behavior being much more robust for this orientation. In fact,
from the field dependence of $T_{FL}$ for $H\perp c$ displayed in Fig.~\ref{HTphasediagram} we can
roughly estimate the metamagnetic transition field $H_M$ to be between 70 T and 90 T for $H\perp
c$.

In conclusion, specific heat and resistivity measurements of \Ir\ in magnetic field up to 17 T
revealed NFL behavior in both of these properties for the field out of the plane ($H \parallel c$,
easy axis) orientation. This behavior develops above 12 T. Fermi liquid is robust for $H\perp c$ in
this field range. On the basis of the phase diagram and the entropy analysis we suggest that the
NFL behavior in \Ir\ is due to the metamagnetic QCP point.

We thank R. G. Goodrich, C. Varma, and J. Lawrence for stimulating discussions. Work at Los Alamos
National Laboratory was performed under the auspices of the U.S. Department of Energy. One of us
(FR) thanks Reines Postdoctoral Fellowship (DOE-LANL) for support. Work at the NHMFL was performed
under the auspices of the National Science Foundation, the State of Florida and the US Department
of Energy.


\end{document}